# PROGRESS TOWARDS DOUBLING THE BEAM POWER AT FERMILAB'S ACCELERATOR COMPLEX

I. Kourbanis[#], FERMILAB*, Batavia, IL 60510, U.S.A.

*Abstract*

After a 14 month shutdown accelerator modifications and upgrades are in place to allow us doubling of the Main Injector beam power. We will discuss the past MI high power operation and the current progress towards doubling the power.

## PAST MI HIGH POWER OPERATION

In the Tevatron era MI operated in a mixed mode providing beams to both the antiproton target and to NuMI neutrino target. Slip stacking [1] was used in order to increase the intensity.

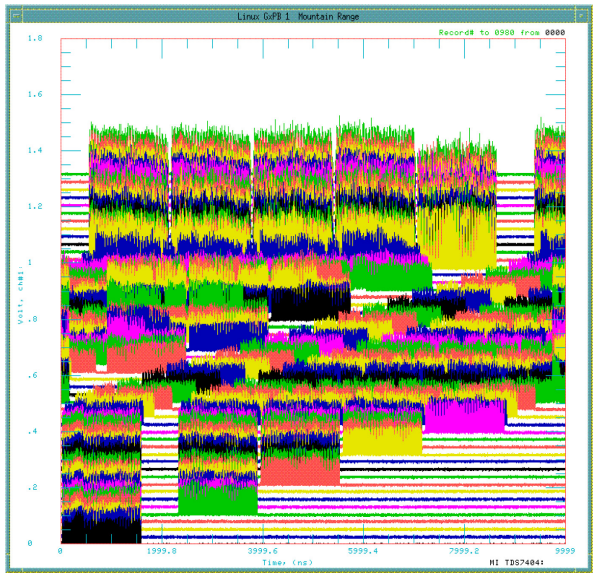

Figure 1: Mountain range of MI mixed mode operation in the TeV era. Two Booster slip stacked batches were delivered to the antiproton target and 9 batches (8 slip stacked plus one single) to the NuMI target.

The overall efficiency was 95% with most of the losses from un-captured beam. A series of MI upgrades were implemented in order to control the losses [2] before we could reliably run in this mode and increase the MI beam power. The typical flattop beam intensity per cycle was 42E12 and the cycle time was 2.2 sec with 0.8sec spent at injection energy for stacking.

We were able to run with peak beam power of 360-380 KW from 2009-2012. Typically around 75 KW were delivered to the antiproton target and 290 KW to the NuMI target. The operating efficiency was around 72%. Most of the downtimes to the neutrino beam were due to target failures and were not related to the accelerator. During the NuMI downtimes the MI continue to deliver beam to the antiproton target. From October 1 2009 to August 2010 (a period with no target failures) a total of 30E19 protons were delivered to the NuMI target and 7E19 protons to the antiproton target. After TEV operations were terminated MI continued to provide high intensity beam to the NuMI power till the ANU shutdown. A total of 9 Booster batches (8+1) per pulse were delivered to the NuMI target with a 2.03 sec rep rate corresponding to 350 KW of beam power.

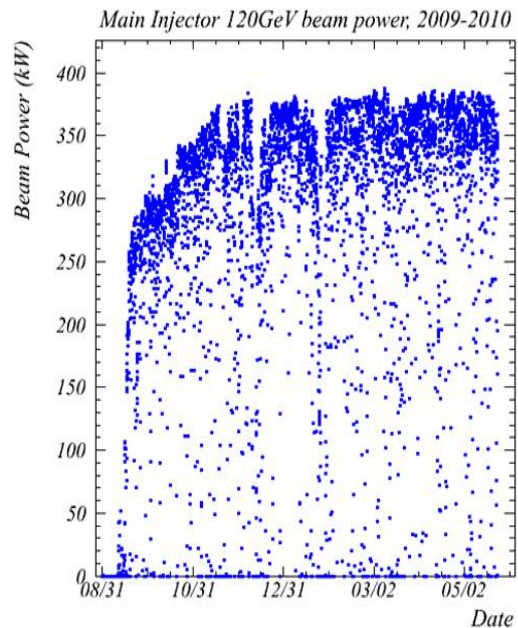

Figure 2: MI beam power from Aug. 2009 to July 2010. The points correspond to the average MI power over an hour.

## PLAN FOR DOUBLING THE MI BEAM POWER

The key element in increasing the MI beam power is the elimination of the long dwell time at 8 GeV. This is accomplished by utilizing the Recycler for injecting and stacking the protons from Booster. The required upgrades were executed as part of ANU (Accelerator and NuMI Upgrades) [3] which was a part of the NOvA project.

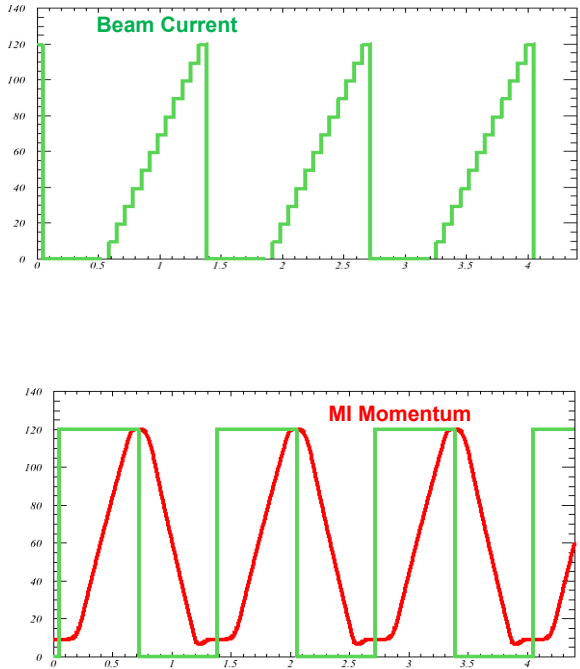

Figure 3: MI and Recycler operation for NOvA. 12 Booster batches are injected and slipped stacked in Recycle while MI is accelerating (top). Beam is injecting in MI from Recycler in one turn and then is accelerated (bottom).

Under the new plan the MI cycle time is reduced from 2.2 sec (33 Booster ticks) to 1.33 sec (20 Booster ticks). The total beam intensity at flattop will be increased to 49E12 (12 Booster batches instead of 11). Up to 8 additional Booster batches can be injected in Recycler for delivery to the modified antiproton rings (Mu2e-g-2 experiments). The high power MI operation requires 9 Hz out of the Fermilab's Booster. The new configuration of the Fermilab's accelerator complex is shown in Fig. 4.

## RECYCLER COMMISSIONING

ANU only provided us with the capability to transform Recycler into a high intensity proton machine with slip stacking. Extensive commissioning was required in order to integrate Recycler into operations. During the Recycler commissioning we operated the MI with 6 batches from the Booster and a 1.67 cycle time providing 280 KW of beam power to the NuMI target.

After establishing slip stacking an extensive period of beam scrubbing was followed in order to increase the Recycler intensity. The Recycler vacuum system [4] is based on TSPs (Ti sublimation pumps) so the beam scrubbing had to proceed slowly. Finally we had to re-configure the MI loss monitors to integrate on Recycler losses before we were able to establish a $2A cycle where we could routinely run the Recycler under the MI cycle (Fig. 5).

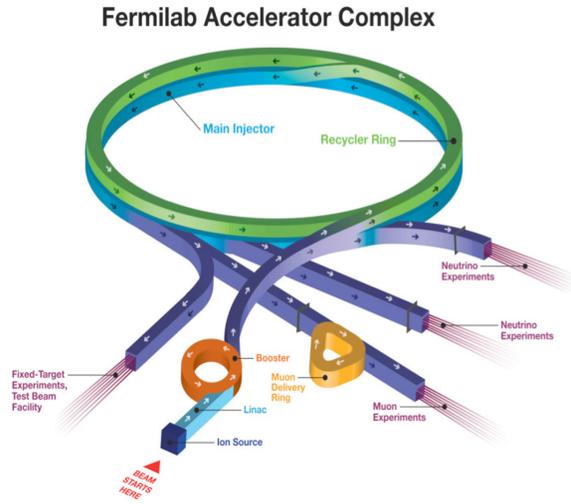

Figure 4: New configuration of the Fermilab's Accelerator Complex

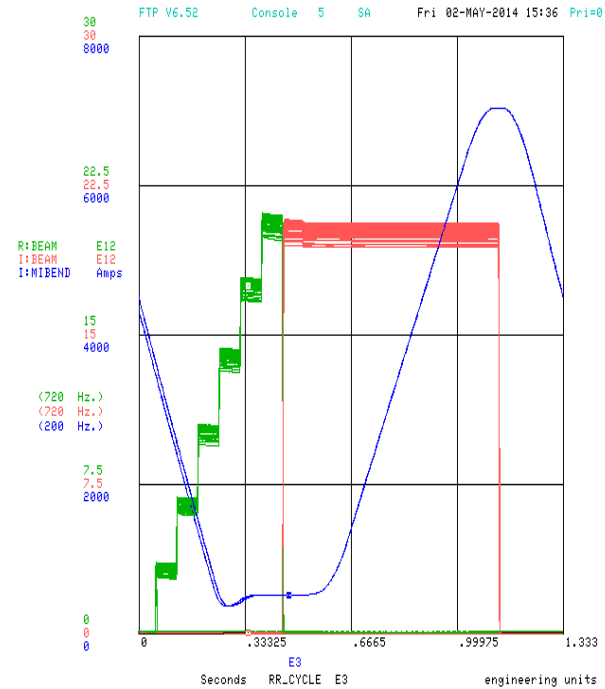

Figure 5: Six Booster batches (green trace) are injected in Recycler during the MI down ramp (blue trace), and then the total beam is injected in MI (red trace).

## CURRENT STATUS AND PLANS

With Recycler operational the MI high intensity operations have switched from the $23 cycles (beam only to MI) to the $2A cycles (stacking in Recycler and then transferring in MI). Initially we continue to use 6 injections from the Booster per cycle. The comparison in beam power between the $23 and all $2A cycles is shown in Figure 6.

Because of the difference in cycle times between the $23 and $2As (1.67 sec vs 1.33 sec) we can get 25% more beam power from MI for the same beam intensity.

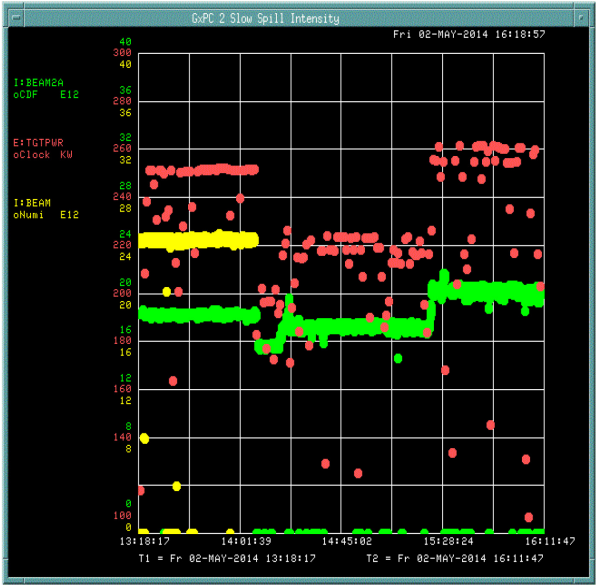

Figure 6: MI beam power (red dots) for two different cycles; $23s (intensity per cycle in yellow) and $2As (intensity per cycle in green).

A typical picture of our current operations is shown in Figure 7. All MI cycles in a 60.0 sec accelerator super cycle are shown ( $2A) along with their typical intensities and the NuMI beam power. Running with a 6 sec switchyard slow spill cycle per super cycle reduces the MI NuMI power by 10%.

We plan to increase the intensity in the $2A cycles to 25E12 from 22E12 increasing the beam power to 350 KW (six batches no slip stacking). To further increase the beam intensity and power we will increase the number of Booster batches to 8 (4 slipped stacked plus 4 single). This will raise our beam power to 460 KW. Running with 8 Booster batches every 1.33 sec requires 6 Hz beam operation out of the Booster which is close to the limit that we can reliably run. Running the Booster at a higher rate will require the completion of the RF cavities refurbishment which is expected to finish in another year.

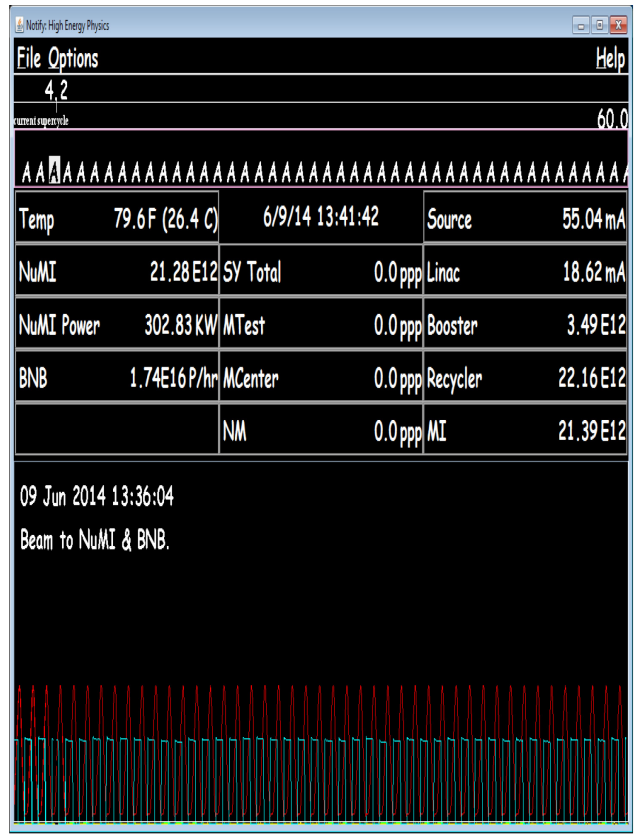

Figure 7: Typical picture of current accelerator operations. The NuMI power is 302.8 KW with 22E12 p in the Recycler.

## CONCLUSIONS

With the accelerator upgrades and the initial commissioning complete Fermilab's Recycler is now operational as a proton injector allowing MI to run with 1.33sec rep. rate. The initial beam power delivered to NuMI target is 300 KW and it expected to increase to 460 KW by the end of the summer. We expect to reach 700 KW when the Booster RF upgrades are complete.